\documentclass[conference]{IEEEtran}

\usepackage{amsmath}
\usepackage{graphicx}
\usepackage[hidelinks]{hyperref}
\usepackage{algpseudocode}
\usepackage{algorithm}
\usepackage{xcolor}
\usepackage{ulem}
\usepackage{multirow}
\usepackage{tikz}
\usepackage{hyperref} 

\begin{document}

\title{Running hardware-aware neural architecture search on embedded devices under 512MB of RAM\vspace{-10pt}}

\author{\IEEEauthorblockN{Andrea~Mattia~Garavagno\IEEEauthorrefmark{1}\IEEEauthorrefmark{2},
        Edoardo~Ragusa\IEEEauthorrefmark{1},
        Paolo~Gastaldo\IEEEauthorrefmark{1},
        and~Antonio~Frisoli\IEEEauthorrefmark{2}
        \IEEEauthorblockA{\IEEEauthorrefmark{1} Department of Electrical, Electronic, Telecommunication Engineering and Naval Architecture, DITEN \\ University of Genoa, Genoa 16145, Italy}
        \IEEEauthorblockA{\IEEEauthorrefmark{2} Department of Excellence in Robotics \& AI, Institute of Mechanical Intelligence, \\ Scuola Superiore Sant'Anna, Pisa 56127, Italy \\ e-mail: AndreaMattia.Garavagno@\{edu.unige.it, santannapisa.it\}}}
        \vspace{-20pt}
}
%
%

\IEEEoverridecommandlockouts

\maketitle

\begin{tikzpicture}[remember picture,overlay]
\node[anchor=south,yshift=15pt] at (current page.south) {
    \parbox{\textwidth}{
        \centering \scriptsize
        This work has been accepted for publication in the proceedings of the 2024 IEEE International Conference on Consumer Electronics (ICCE). \copyright~2024 IEEE. Personal use of this material is permitted.  Permission from IEEE must be obtained for all other uses, in any current or future media, including reprinting/republishing this material for advertising or promotional purposes, creating new collective works, for resale or redistribution to servers or lists, or reuse of any copyrighted component of this work in other works. Final publication is available at \url{https://doi.org/10.1109/ICCE59016.2024.10444268}.
    }
};
\end{tikzpicture}

\IEEEpubidadjcol

\begin{abstract}
This document proposes a novel approach to hardware-aware neural architecture search (HW NAS) that considers the resources available on the computing platform running it, enabling its execution on various embedded devices. The presented HW NAS produces tiny convolutional neural networks (CNNs) targeting low-end microcontroller units (MCUs), typically involved in the Internet of Things (IoT) or wearable robotics, opening new use cases. A gateway could run it to tailor CNNs' architecture on the acquired data without using external servers, ensuring privacy. The proposed technique achieves state-of-the-art results in the human-recognition tasks on the Visual Wake Word dataset, a standard TinyML benchmark, on several embedded devices.
\end{abstract}

\begin{IEEEkeywords}
TinyML, Hardware Aware Neural Architecture Search, Internet of Things, Wearable Robotics  
\vspace{-10pt}
\end{IEEEkeywords}

\IEEEpeerreviewmaketitle

\section{Introduction}
\vspace{-5pt}
Hardware-aware neural architecture search (HW NAS) automates the design of neural architectures considering hardware constraints imposed by the target device that support the prediction phase of the generated model.

The first HW NASs exploited reinforcement learning (RL) to search for the best architecture: a controller produced instances of architectures and was rewarded based on the accuracy obtained after the training and the hardware cost \cite{MNASnet,FPNet}. The search cost was high both in terms of FLOPS and memory requirements. MNASNet \cite{MNASnet} required 40,000 GPU hours of execution to obtain the presented results. 

Over-parametrized networks, known as supernetwork, containing all the models belonging to the search space as sub-networks reduced significantly search time \cite{DARTS}. Such a solution allowed training only one big network and then searching inside it for the best sub-network. Techniques like evolutionary algorithms \cite{MCUNet,ONCEFORALL} and gradient descent methods \cite{DARTS,Micronets} were commonly used to search for the best sub-network. Even though the search time had been reduced, a super network still required high-end GPUs to be trained. MCUNet \cite{MCUNet} required 300 GPU hours of execution to obtain the presented results. 

Bi-level optimization methods allowed to further reduce search costs down to 20 GPU hours \cite{fast_and_practical} and eventually remove the need for a GPU \cite{garavagno_PRIME23}, obtaining the resulting architecture in 3.5 hours of execution on a laptop. 

The present research introduces a novel approach to HW NAS. Besides the constraints given by the target hardware, it also considers the resources available on the computing platform running it, enabling its execution on embedded devices. 
\vspace{-17pt}

\section{Proposal}
The present work proposes a novel approach to HW NAS which considers the resources available on the computing platform running it, besides the constraints given by the target hardware. To implement, an existing HW NAS \cite{garavagno_PRIME23} producing tiny convolutional neural networks (CNNs) for low-end microcontroller units (MCUs) has been modified. 
Three main design choices characterize HW NAS procedures: the search space, the formulation of the optimization problem, and the search strategy. The search space sets the admissible network architectures. The optimization problem establishes the search boundaries and the evaluation metric used. Finally, the search strategy describes how the search space is explored, i.e., how the solution is found.


The search space contains candidate solutions based on stacks of cells with a fixed structure. Each cell has three layers: a max pooling layer followed by a batch normalization layer which feeds a convolutional layer. The max pooling layer always halves the input resolution. Convolutional layers use a fixed kernel size of (3,3) with stride 1. 

Cells are stacked upon a first convolutional layer having kernels, preceded by a fixed pre-processing pipeline performing min-max standardization and batch normalization on the input data. A global average pooling layer with dropout regularization feeds a final layer with softmax activation for classification. The number of kernels used in each cell was set using the heuristic proposed in \cite{garavagno_PRIME23} that sets the number of kernels as a function of the selected value $k$ for the first convolutional layer and as a function of the number of cells.

Eventually, a candidate architecture has two parameters: $k$ and $c$. $c$ defines the number of cells stacked. This tiny search space includes a few thousand possible architectures, which can even lessen in cases of really tight resource constraints.

Equation \ref{eq:problem} shows the new optimization problem proposed in the present paper, where function $f$ returns the maximum validation accuracy after three epochs of training. The constraints $\phi_{R}$, $\phi_{F}$, $\phi_{M}$, represent respectively peak RAM occupancy, Flash occupancy, and number of MACC. $\phi_{E}$ represents the available RAM of the embedded device running the HW NAS. Thanks to the novel constraint, during the search space exploration, candidate solutions that do not fit the embedded device running the HW NAS but that would fit the target hardware are discarded to allow the execution of the algorithm on the embedded platform. 
\vspace{-15pt}

\begin{equation} \label{eq:problem}
\begin{aligned}
P: \begin{cases}
\max f(x)\\
   \phi_{E}(x) \leq \xi_{E} \\
   \phi_{R}(x) \leq \xi_{R} \phi_{F}(x) \leq \xi_{F}, \phi_{M}(x) \leq \xi_{M}    \\
   \xi_{R}, \xi_{F}, \xi_{M}, \xi_{E} > 0 
\end{cases}
\end{aligned}
\end{equation}

The optimization problem $P$ is solved with a derivative-free optimization technique proposed in \cite{garavagno_PRIME23}.

\section{Experimental Results}
The implemented HW NAS has been run on four different embedded platforms: the NVIDIA® Jetson Nano™ with 4 GB of RAM, operated in the MAXN power mode (J. Nano), the Raspberry Pi 4 Model B with 4 GB of RAM (RPi 4), the Raspberry Pi 3 Model B (RPi 3), and the Raspberry Pi Zero 2 W (RPiZ 2 W). All the boards mounted a passive heatsink. ZRAM kernel module was used with the lzo-rle compression algorithm and a swap area equivalent to 2/3 of the available memory for all the platforms. When possible, the dataset was cached in RAM.

The Visual Wake Words dataset, a standard tinyML benchmark \cite{TinyML_benchmarks}, has been adopted. To have a comparison with the state-of-the-art, the largest among the smallest deployment targets, in terms of RAM occupancy, of two famous HW NAS MCUNET and Micronets has been chosen: the STM32F412 MCU. The latter is to balance the fact that they focused on STMicroelectronics' high-performance MCUs, while the proposed work focused on low-end MCUs. Input size was set at (50,50,3).

Table \ref{tab:results} shows the test accuracy, RAM and Flash occupancy, multiply and accumulate (MAC) instructions, and search costs for each model obtained running the proposed HW NAS on the corresponding embedded platform using the STM32F412 as a deployment target and the Visual Wake Words dataset as a benchmark. All the models are saved in the TFLite format.

\begin{table}[!ht]
        \centering
        \begin{tabular}{c | c c c c c} 
                        Embedded  & Acc  & RAM   & Flash & MACC & Time   \tabularnewline 
                        Platform  & [\%] & [kiB] & [kiB] & [MM] & [dd]   \tabularnewline \hline
                        J. Nano   & 78   & 36    & 38.96 & 2.6  & 3.9  \tabularnewline 
                        RPi 4     & 77.6 & 33.50 & 32.84 & 2.07 & 1.3  \tabularnewline 
                        RPi 3     & 76.6 & 28.5  & 23.65 & 1.3  & 3.8  \tabularnewline 
                        RPiZ 2 W  & 75.1 & 25    & 13.95 & 0.91 & 4.6 \tabularnewline 
                                  &      &       &       &      &         
        \end{tabular} \\ 
        \caption{\label{tab:results} Test accuracy, RAM and Flash occupancy, the number of MAC instructions, search costs for each model obtained running the proposed HW NAS on the corresponding embedded platform using the STM32F412 as a deployment target and the Visual Wake Words dataset as a benchmark. \vspace{-15pt}}
\end{table}

Table \ref{tab:state_of_the_art} shows for reference purposes the test accuracy, RAM and Flash occupancy, the number of MAC instructions, and hardware target declared by Micronets \cite{Micronets} and MCUNET \cite{MCUNet} for the Visual Wake Words dataset. Also these models are saved in the TFLite format.

\begin{table}[!ht]
        \centering
        \begin{tabular}{c | c c c c c c c c} 
                        \multirow{2}{*}{Project}           & Acc  & RAM   & Flash  & MACC & input      &  target  \tabularnewline 
                                                           & [\%] & [kiB] & [kiB]  & [MM] & [w,h,c] & [STM32]  \tabularnewline \hline
                        \href{https://hanlab.mit.edu/projects/tinyml/mcunet/release/mcunet-10fps_vww.tflite}{\color{blue}{MCUNet}}       & 87.4 & 168.5 & 530.52 & 6    & 64,64,3  & F412     \tabularnewline 
                        \href{https://github.com/ARM-software/ML-zoo/tree/master/models/visual_wake_words/micronet_vww2/tflite_int8}{\color{blue}{Micronets}} & 76.8 & 70.5  & 273.81 & 3.3  & 50,50,1  & F446RE   \tabularnewline 
                                                           &      &       &        &      &         &    
        \end{tabular} \\ 
        \caption{\label{tab:state_of_the_art} Test accuracy, RAM and Flash occupancy, the number of MAC instructions, and resolution declared by Micronets and MCUNET for each target for the Visual Wake Words dataset. \vspace{-22pt}}
\end{table}

The obtained models are remarkably smaller than the ones from MCUNet and Micronets while having an accuracy comparable to Micronets. The resulting search costs are in the order of days because of the nature of the devices involved in the experiment, which are not meant to run HW NAS. A proxy dataset of 10k images could be used to make the search cost more practical. The execution times do not decrease linearly w.r.t. the increase in computational capabilities because, in this case, it is associated with an increase in memory, which allows more candidate solutions to be explored. Noteworthy is the RPi 4 execution time, which is significantly lower than the one of J. Nano because of the absence of the GPU, which left more free memory, allowing the cache of the whole dataset in memory.
\vspace{-10pt}

\section{Conclusion}
The present research introduced a novel approach to HW NAS, which also considers the resources available on the computing platform running it, enabling its execution on embedded devices. Its implementation produced an HW NAS meant to design tiny CNNs targeting low-end MCUs, obtaining state-of-the-art results on several embedded devices. Being able to automatically design tiny CNNs for low-end MCUs, typically involved in the IoT or wearable robotics, on embedded devices opens new use cases. For example, a gateway node could run the proposed HW NAS to tailor CNNs' architecture on the acquired data without using external servers, ensuring privacy.

\ifCLASSOPTIONcaptionsoff
  \newpage
\fi



\bibliographystyle{IEEEtran}
\bibliography{bibtex/bib/IEEEabrv, bibtex/bib/refs}
%

%




\end{document}